\documentclass{article}
\usepackage{epsfig}
\usepackage{amssymb}

\newcommand{\bc}{\begin{center}}
\newcommand{\ec}{\end{center}}
\newcommand{\bmi}{\begin{minipage}}
\newcommand{\emi}{\end{minipage}}
\newcommand{\bdi}{\begin{displaymath}}
\newcommand{\edi}{\end{displaymath}}
\newcommand{\bit}{\begin{itemize}}
\newcommand{\eit}{\end{itemize}}

\newcommand{\Ne}{N\'{e}el}

\newcommand{\bi}{\begin{itemize}}
\newcommand{\ei}{\end{itemize}}
\newcommand{\be}{\begin{equation}}
\newcommand{\ee}{\end{equation}}

\newcommand{\bea}{\begin{eqnarray}}
\newcommand{\eea}{\end{eqnarray}}   

\newcommand{\bdm}{\begin{displaymath}}
\newcommand{\edm}{\end{displaymath}}
\newcommand{\beas}{\begin{eqnarray*}} 
\newcommand{\eeas}{\end{eqnarray*}}
\newcommand{\bite}{\begin{itemize}}
\newcommand{\enite}{\end{itemize}}
\newcommand{\ben}{\begin{enumerate}}
\newcommand{\een}{\end{enumerate}}

\begin{document}

\title{Quantum phase transition in the Plaquette lattice 
with anisotropic spin exchange}

\author{A. Voigt
\footnote{e-mail: andreas@physast.uga.edu, Tel. +1-706-542-3884, 
Fax +1-706-542-2492},\\
Center for Simulational Physics\\
Department of Physics and Astronomy\\
University of Georgia, Athens GA 30605, USA}

\maketitle

\begin{abstract} 
I study the influence of anisotropic spin exchange on a quantum phase
transition in the Plaquette lattice driven by the purely quantum effect of
singlet formation. I study the influence of i) a Dzyaloshinskii-Moriya
exchange and ii) four spin exchange on the transition point by evaluating
spin--spin correlations and the spin gap with exact diagonalization. The
results point to a stabilization of the \Ne-like long range order when the
Dzyaloshinskii-Moriya exchange is added, whereas the four-spin exchange might
stabilize the singlet order as well as the \Ne-like order depending on its
strength.
\end{abstract}

\vspace*{0.3cm}

{\it quantum antiferromagnets, Plaquette lattice, 
anisotropic spin exchange}


\section{Introduction}

Low dimensional quantum antiferromagnets show show a wide variety of
magnetic low--temperature behavior, like magnetic long range order or spin
disorder with or without a spin gap. Based on the findings of a recent
experimental study of a new compound $Na_5RbCu_4(AsO_4)_4Cl_2$
\cite{clayhold01} I will examine a quantum spin system which shows a purely
quantum phase transition by singlet formation. This and related models have
been studied previously \cite{koga99,singh99,voigt2002}. From this results
one argues that the observed low-temperature magnetic transition in
$Na_5RbCu_4(AsO_4)_4Cl_2$ can probably not be explained within a simple
Heisenberg-type model approach. Therefore additional anisotropic spin
interactions have to be taken into account and in this paper I will study
the influence of a Dzyaloshinskii-Moriya (DM) interaction and a four-spin
exchange interaction on the quantum phase transition in the Plaquette
lattice.

\section{The Model} 

I consider an antiferromagnet on a two--dimensional
square lattice with different types of interactions:
\bea
H & = &{J_p} \sum_{\Box}{{\hat s}_i{\hat s}_j} +
   {J_n} \sum_{\not\subset \Box}{{\hat s}_i{\hat s}_j} \nonumber \\
  & + & \sum_{\Box} {\bf D}_{ij} ({\hat s}_i \times {\hat s}_j) +
        {W_{4S}} \sum_{\Box} ({\hat s}_i {\hat s}_j 
        {\hat s}_k {\hat s}_l) 
\eea
The interactions are: $J_p$ -- Heisenberg type between 4 spins building a
plaquette $\Box$, $J_n$ -- Heisenberg type between plaquettes on a simple
square lattice, ${\bf D}_{ij}$ -- DM-type between plaquette spins and 
$W_{4S}$ -- a four-spin interaction between plaquette spin respectively.

\begin{figure}[ht]
\bc
\epsfig{file=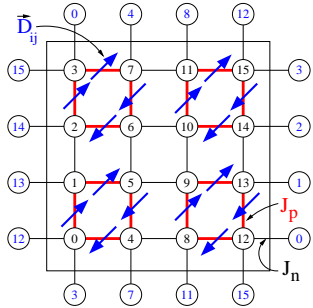,scale=0.28,angle=0}
\ec
\caption{
The N=16 lattice with periodic boundary conditions and interactions: 
$J_p$, $J_n$ and ${\bf D}_{ij}$.}
\label{latt}
\end{figure}

The DM interaction ${\bf D}_{ij}$ is a vector the components of which have
to be extracted from the lattice symmetries \cite{voigt96jpc}. Here we
consider one particular realization ${\bf D}_{ij}=(W_{DM},W_{DM},0)$ where
the x- and y-components are equal. Due to symmetry considerations the
interaction reverse sign on adjacent plaquette bonds (see Fig.\ref{latt}).
The four-spin exchange $W_{4S}$ emerges from the strong-coupling expansion
of the Hubbard model and has been recently discussed in ladders and
two-dimensional lattices \cite{matsuda2000,coldea2001}.

\section{Results} 

It has been shown that without anisotropic exchange (i.e. $W_{DM}=0$ and
$W_{4S}=0$) there is a critical $J_n \approx 0.55$ where the Plaquette
lattice changes from a disordered spin gap state to a long-range ordered
\Ne-like ground state \cite{voigt2002}. I calculate for small finite
lattices (N=16,20) the quantum ground and first excited states with exact
diagonalization. 

\begin{figure}[ht]
\bc
\epsfig{file=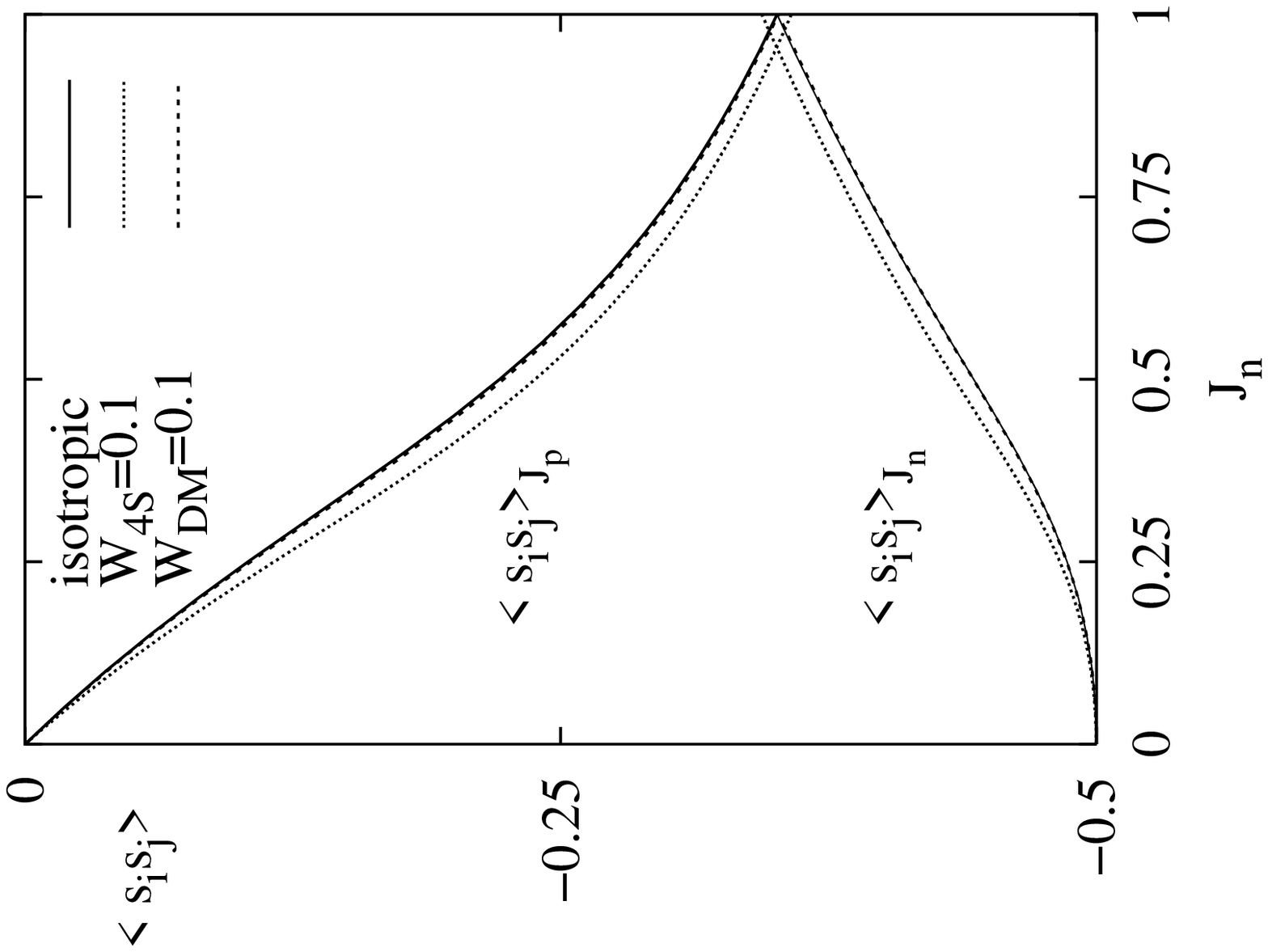,scale=0.25,angle=270}
\epsfig{file=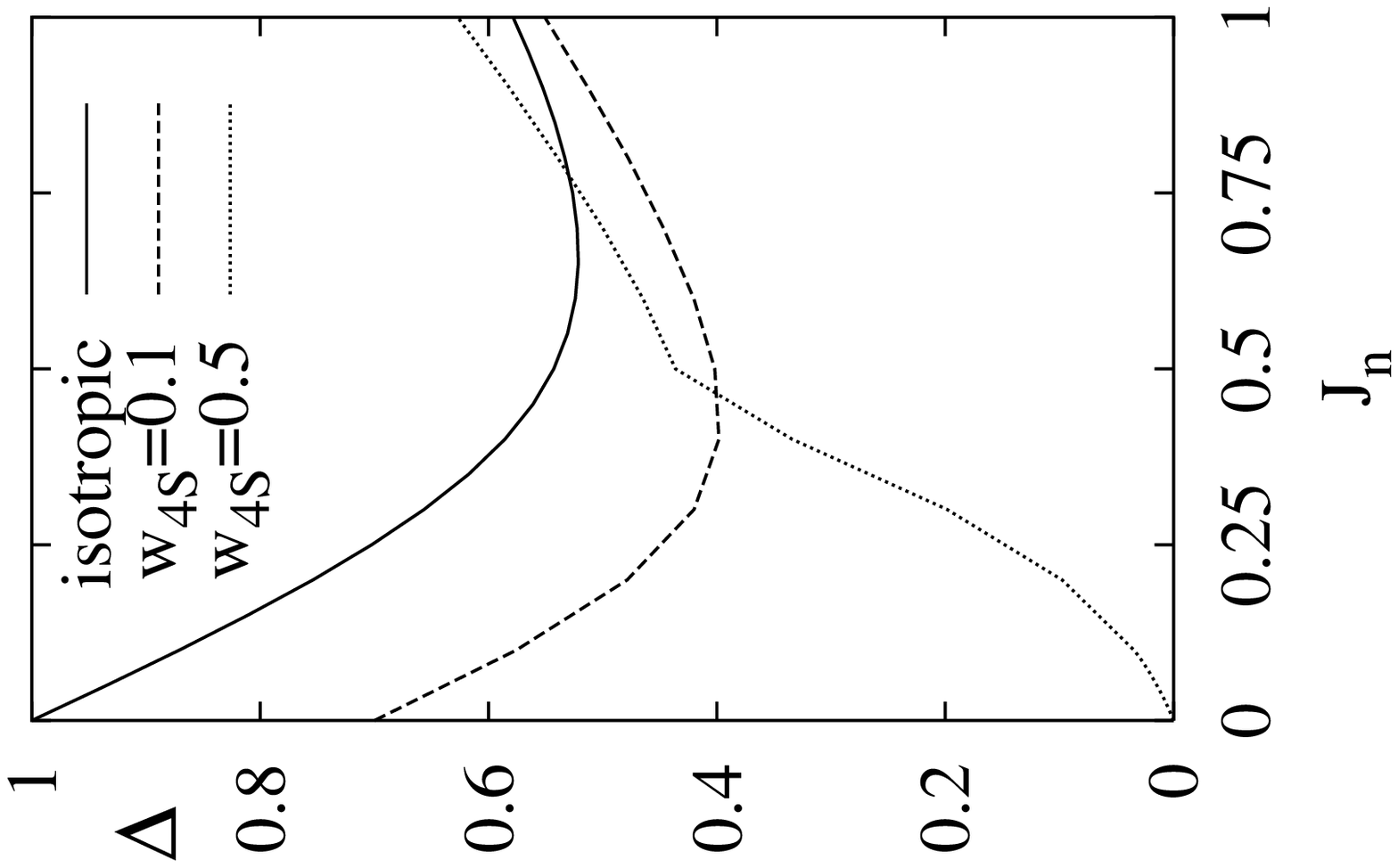,scale=0.25,angle=270}
\ec
\caption{
Selected data of the spin--spin correlation ${\langle \hat s_i \hat s_j
\rangle}$ on a $J_p$ or an $J_n$ bond (left) and for the
spin gap $\Delta$ )(right) for different values of $W_{DM}$ and $W_{4S}$ 
for N=16.}
\label{data}
\end{figure}

By analyzing the differential spin correlations between plaquettes and the
spin gap for selected values of $W_{DM}$ and $W_{4S}$ I extract the critical
value of $J_n$ where the phase transition takes place.

\begin{figure}[ht]
\bc
\epsfig{file=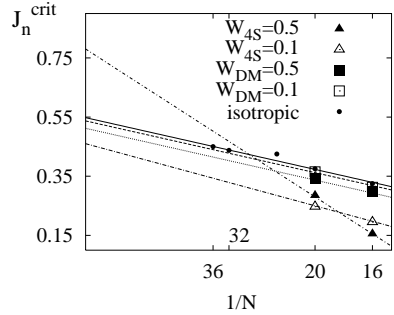,scale=0.3,angle=270}
\ec
\caption{
The extracted values for $J_n^{crit}$ for
different values of $W_{DM}$ and   
$W_{4S}$ and different lattices sizes.}
\label{extr}
\end{figure}

For the particular chosen DM interaction we see a shift of $J_n^{crit}$ to
smaller values. We argue that this type of DM interaction always
destabilizes the plaquette order by disturbing the singlet building on them.
Interestingly enough, a related DM interaction on the $J_1$--$J_2$ square
lattice studied previously \cite{voigt96jpc} also showed a similar effect.

For small values of $W_{4S}$ we observe a similar behavior as in the case of
the DM interaction. But interestingly for larger $W_{4S}$ the tendency to
smaller $J_n^{crit}$ reverses and it seems that at some strength of $W_{4S}$
the corresponding $J_n^{crit}$ becomes larger then for the isotropic system.
That means that the four-spin exchange $W_{4S}$ can have two opposite effect
depending in its strength. It either stabilizes the \Ne-like order on the
lattice or (if strong enough) can help to build singlets on the plaquettes.

Additional calculations on larger lattices will be carried out in order to
confirm these conclusions. 

There are several open questions; about the influence of the particular
chosen symmetry of the DM interaction, about the result of a mixing of
both anisotropic interactions (which might reveal new and interesting order
phenomena) and about the nature of the phase transition (first or second
order). In a more detailed study I will consider those and other problems 
and report the results elsewhere.

This work was supported by NSF grant ACI-0081789.

\end{document}